\documentclass[12pt]{article}

\usepackage{graphicx}
\usepackage{amsfonts}
\usepackage{amssymb}
\usepackage{epstopdf}
\usepackage{multirow}

\usepackage[T1]{fontenc}
\usepackage{bbold}

\let\ssection=\section
\renewcommand{\section}{\setcounter{equation}{0}\ssection}

\newcommand\mathC{\mkern1mu\raise2.2pt\hbox{$\scriptscriptstyle|$}
        {\mkern-7mu\rm C}}              
\newcommand{\mathR}{{\rm I\! R}}         

\newcommand\be{\begin{equation}}
\newcommand\ee{\end{equation}}
\newcommand{\raw}{\rightarrow}

\def\a{\alpha}

\def\d{\delta}

\def\dd{\mbox{d}}

\def\bra{\langle}
\def\ket{\rangle}
\def\a{\alpha}

\def\d{\delta}
\def\D{\Delta}

\def\f{\phi}

\def\m{\mu}

\def\pa{\partial}

\newcommand{\tn}[1]{\mbox{\tiny #1}}
\renewcommand{\@}[1]{\sqrt{#1}}
\renewcommand{\le}[1]{\label{#1}\end{eqnarray}}
\newcommand{\bea}{\begin{eqnarray}}
\newcommand{\eea}{\end{eqnarray}}

\newcommand{\eq}[1]{(\ref{#1})}

\def\ffract#1#2{\raise .35 em\hbox{$\scriptstyle#1$}\kern-.25em/
\kern-.2em\lower .22 em \hbox{$\scriptstyle#2$}}
\def\GN{G_{\mbox{\tn N}}}

\def\na{\nabla}

\begin{document}

\pagestyle{empty}

\centerline{{\Large \bf Comparing Dualities and Gauge Symmetries}}
\vskip 1.5truecm

\begin{center}
{\large Sebastian De Haro$^{1,2}$, Nicholas Teh$^{1,3}$, Jeremy N.~Butterfield$^4$}\\
\vskip .7truecm
{\it $^1$Department of History and Philosophy of Science, University of Cambridge\\
Free School Lane, Cambridge CB2 3RH, United Kingdom}\\
{\it $^2$Amsterdam University College, University of Amsterdam,
Science Park 113\\ 1090 GD Amsterdam, The Netherlands}\\
{\it $^3$ Department of Philosophy, University of Notre Dame \\
South Bend, IN, USA} \\
{\it $^4$ Trinity College, Cambridge, CB2 1TQ, United Kingdom}

\vskip .3truecm
{\tt sd696@cam.ac.uk, njywt2@cam.ac.uk, jb56@cam.ac.uk}\\
\vskip.1truecm
27 March 2016
\end{center}
\vskip .3truecm

\begin{center}

\textbf{\large \bf Abstract}
\end{center}
We discuss some aspects of the relation between dualities and gauge symmetries. Both of these ideas are of course multi-faceted, and we confine ourselves to making two points. Both points are about dualities in string theory, and both have the `flavour' that two dual theories are `closer in content' than you might think. For both points, we adopt a simple conception of a duality as an `isomorphism' between theories: more precisely, as appropriate bijections between the two theories' sets of states and sets of quantities. 
 
The first point (Section \ref{thydlty}) is that this conception of duality meshes with two dual theories being `gauge related' in the general philosophical sense of being physically equivalent. For a string duality, such as T-duality and gauge/gravity duality, this means taking such features as the radius of a compact dimension, and the dimensionality of spacetime, to be `gauge'.

The second point (Sections 4, 5 and 6) is much more specific. We give a result about gauge/gravity duality that shows its relation to gauge symmetries (in the physical sense of symmetry  transformations that are spacetime-dependent) to be subtler than you might expect. For gauge theories, you might expect  that the duality bijections relate only gauge-invariant quantities and states, in the sense that gauge symmetries in one theory will be unrelated to any symmetries in the other theory.  This may be so in general; and indeed, it is suggested by discussions of Polchinski and Horowitz. But we show that in gauge/gravity duality, each of a certain class of gauge symmetries in the gravity/bulk theory, viz.~diffeomorphisms, {\em is} related by the duality to a position-dependent symmetry of the gauge/boundary theory.

\newpage
\pagestyle{plain}

\tableofcontents

\newpage

\section{Introduction}\label{intro}

The ideas of duality and gauge symmetry are among the most prominent in modern physics; and they are both related to the philosophical questions of how best to define (a) physical theories, and (b) the relation of theoretical equivalence (of `saying the same thing') between theories. In this paper, we will make two main points about how duality and gauge symmetry are connected.

Both points are about dualities in string theory, and both have the `flavour' that two dual theories are `closer in content' than you might think. For both points, we adopt a simple conception of a duality as an `isomorphism' between theories. In Section \ref{thydlty}, we state this conception, and briefly relate it to the philosophical questions (a) and (b). In short, we take a theory to be given by a triple comprising a set of states, a set of quantities and a dynamics; so that a duality is an appropriate `structure-preserving' map between such triples. This discussion will be enough to establish our first point. Namely: dual theories can indeed `say the same thing in different words' ---which is reminiscent of gauge symmetries. 

Our second point (Sections 4, 5 and 6) is much more specific. We give a result about a specific (complex and fascinating!) duality in string theory, gauge-gravity duality: which we introduce in Section \ref{ggd}, using Section \ref{thydlty}'s conception of duality. We state this result in Section \ref{invarces}. (Details, and the proof, are in De Haro (2016).) It says, roughly speaking, that each of an important class of  gauge symmetries in one of the dual theories (a gravity theory defined on a bulk volume) is mapped by the duality to a gauge symmetry of the other theory (a conformal field theory defined on the boundary of the bulk volume). This is worth stressing since some discussions suggest that all the gauge symmetries in the bulk theory will not map across to the boundary theory, but instead be `invisible' to it. As we will see, this result also prompts a comparison with the hole argument. It also relates to recent discussion of the empirical significance of gauge symmetries---a topic we take up in Section \ref{gauge}.	 

To set the stage for these points, especially the second, Section 2 will describe the basic similarity between the ideas of duality and gauge symmetry: that they both concern `saying the same thing, in different words'.  

\section{`Saying the same thing, in different words'}\label{samedifferent}

We have already sketched the idea of a {\em duality} as a `structure-preserving' map between two physical theories, or between two formulations of a single theory. Typically, there is a map that relates the theories---states, quantities and dynamics---in a `structure-preserving way'. 

As to {\em gauge symmetry}, we need to distinguish:\footnote{We should note that many theoretical physicists would take `gauge symmetries' to mean `small and asymptotically trivial symmetries', at least in the case of a subsystem. However, our labeling will serve as a helpful clarification, since `gauge' is a term about which there is still much confusion.} (i) a general philosophical meaning, and (ii) a specific physical meaning; and similarly, for cognate terms like `gauge theory', `gauge-dependent' and `gauge-invariant'. As follows:---\\
\indent (i) (Redundant):  If a physical theory's formulation is redundant (i.e.~roughly: it uses more variables than the number of degrees of freedom of the system being described), one can often think of this in terms of an equivalence relation, `physical equivalence', on its states; so that gauge-invariant quantities are constant on an equivalence class and gauge-symmetries are maps leaving each class (called a `gauge-orbit') invariant. Leibniz's criticism of Newtonian mechanics provides a putative example: he believed that shifting the entire material contents of the universe by one meter must be regarded as changing only its description, and not its physical state.\\
\indent (ii) (Local):  If a physical theory has a symmetry (i.e.~roughly, a transformation of its variables that preserves its Lagrangian)\footnote{There are two kinds of subtlety, that come up even in the classical context: (i) the relation between preserving the Lagrangian (called a `variational symmetry') and preserving the equations of motion (a `dynamical symmetry'); (ii) the need in Hamiltonian mechanics to preserve the symplectic form, as well as the Hamiltonian. For an exposition of (i) and (ii), cf.~e.g.~Butterfield (2006, especially Section 3.2, 3.4, 5.3, 6.5). Suffice it to say here that, for the  quantum case we will be concerned with in Section \ref{thydlty}f., which will be treated in a Lagrangian i.e.~path-integral framework: a symmetry  must preserve the Lagrangian and the measure.} that transforms some variables in a way dependent on spacetime position (and is thus `local') then this symmetry is called `gauge'. In the context of Yang-Mills theory, these variables are `internal', whereas in the context of General Relativity, they are spacetime variables---both types of examples will occur in Sections 4f. 

Although (Local) is often a special case of (Redundant), it will be important to us that this is not always so. For we will be concerned with (Local) gauge symmetries (specifically: diffeomorphisms) which are asymptotically non-trivial (i.e.~do not tend to the identity at spacelike infinity), and which can thus change the state of a system relative to its environment.\footnote{\label{GreavWall}{Cf.~the discussion in Greaves and Wallace (2014) and  Teh (2015); cf. also Section \ref{gauge}}.}

These sketches are enough to suggest that for any theory, or for any theory and its duals, duality and gauge-symmetry are likely to be related. The obvious suggestion to make is that the differences between two dual theories will be like the differences between two formulations of a gauge theory: they `say the same thing, despite their differences'. Indeed, for several notable dualities, this is the consensus among physicists. But the details vary from case to case, and can be a subtle matter: much depends on how we interpret the vague phrase `will be like'. 

The subtleties arise, in part, from the fact that physicists tend to call a correspondence or map that `preserves what is said' a ``duality'', only when it is: (1) striking and-or (2) useful, in one way or another---and in a way that gauge-symmetries typically are {\em not}.  It is worth spelling out these desiderata a little. It will show how we should interpret the phrase `will be like'; that is, it will enable us to disambiguate and assess the suggestion.

(1): {\em Striking}: One main way the correspondence can be {\em striking} is that the two `sayings' (i.e.~the two formulations) are very disparate. The most striking examples of this occur in string theory (which will be our focus). Thus in S-duality, the two formulations differ about the electric or the magnetic nature of the charges. In T-duality, the two formulations differ about the radius of a compact dimension of space: where one says it is $r$, the other says it is $1/r$. And in gauge-gravity duality (on which we focus from Section \ref{ggd}),  the formulations differ about the dimensionality of spacetime (and of course, much else!): where the `bulk' formulation says it is $d+1$ (say 5), the other `boundary' formulation  says it is $d$ (say 4).

Agreed: such a difference---of electric vs.~magnetic charge, or the radius or dimensionality of spacetime---seems so marked that you might well doubt that the  formulations are `saying the same thing'. You might well say instead that they contradict each other; so that the example is best taken as a case of under-determination (of theory by all possible data), not as any kind of theoretical equivalence. This is a reasonable reaction: for example, McKenzie (2015, Section 4) argues for this view, as regards S-duality. More generally, we agree that the relations between dualities and theoretical equivalence, especially for string theory, are by no means settled, as several recent philosophical  discussions witness.  (Examples for our main case, gauge/gravity duality, which also discuss claims that one side of the duality (usually the gravity side) is `emergent' from the other, include: De Haro (2015, especially Section 3.2), Dieks et al. (2015, especially Section 3.3), Rickles (2012, especially Section 5), Teh (2013, especially Sections 3, 4).) 

 But in this paper, we will not need to decide these issues, nor even the best interpretation of gauge/gravity duality's case of dimensionality; though we will briefly return to the issues in Sections 3 and 4. For the moment, it is enough to report the consensus in string theory: that at least some of these striking dualities {\em do} relate different formulations of a single theory. As the physics jargon has it: they `describe the same physics'; meaning of course not just observational, but also theoretical, equivalence. And several philosophical commentators endorse this consensus, including for our case of gauge/gravity duality; e.g.~De Haro (2015, Section 2.4.1), Dieks et al. (2015,  Section 3.3.2), Huggett (2015, Section 2.1, 2.3), Rickles (2011, Section 2.3, 5.3), Rickles (2012, Section 6), 
Rickles (2015), Matsubara (2013, Section 6).

Besides, some string theorists go further. They take the ongoing search for an M theory to be the search for a formulation which will relate to the present formulations, on the two sides of such dualities, in much the way a gauge-invariant formulation of a gauge theory relates to different choices of gauge.\footnote{Indeed, the discovery of T-duality was one of the factors that prompted the idea of M theory: see Witten (1995).} 

(2): {\em Useful}: One main way the correspondence can be {\em useful} is if it relates a regime where problems are difficult to solve (say because couplings are strong) to one where they are easy to solve (say because couplings are weak). This also happens in gauge/gravity duality: problems that are difficult (because strongly coupled) in the boundary theory are mapped into easier, weakly coupled, problems in the bulk.  (Agreed: a gauge symmetry is often useful for problem solving in that the change of formulation (change of gauge) it implements renders a problem easier to solve. But this advantage usually applies to at most a limited family of problems. For a duality, the advantage usually applies to a whole coupling regime.)

So what is the upshot of these points, (1) and (2), for what we called `the obvious suggestion'? 

\indent  On the one hand, the suggestion is right, in two main respects. First: for any gauge theory, a (Redundant) gauge symmetry preserves `what is said' by the two formulations, i.e.~gauges, that it relates; and although there will usually not be striking differences between these gauges, the gauge change may help solve a problem. To that extent, a gauge symmetry is like a duality. Second: string theorists do usually conceive their duals---be they ever so disparate---as `saying the same physics', in a manner reminiscent of the (much better understood) gauge symmetries.  On the other hand, `duality' tends to be used, in particular in string theory, for a correspondence or map that is more striking and-or useful than a gauge symmetry.

For our purposes, the important point arising from this discussion is that this last contrast tends to foster a {\em false} impression: roughly, that dualities between {\em gauge} theories must ignore (Local) gauge structure. To see how this impression arises, let us for the moment assume (falsely)  that all (Local) gauge symmetries exemplify (Redundant). Then, given a striking and-or useful duality between two theories' gauge-invariant structures, it would be even more striking if the duality also mapped their gauge-dependent structures (gauge symmetries and  gauge-dependent quantities) into each other. After all, think of the everyday analogy in which (i) a duality is like a translation scheme between languages, and so (ii) gauge structure, a theory allowing several gauges, is like a language having several synonyms for one concept. (Indeed, this analogy is entrenched in physics: physicists call the definition of the duality  transformation the `dictionary'  etc.) One would not expect a translation scheme to match the languages' synonym structures, i.e.~to translate each of a set of synonyms in $L_1$ by just one of the corresponding (synonymous!) set of synonyms in $L_2$, and {\em vice versa}. Analogously: it seems that for a duality between gauge theories, the gauge-dependent structures on the two sides will {\em not} be related by the duality. Each such structure, on one side, will be `invisible' to the other side---at least, `invisible' if you are using just the duality map `to look through'.

It is {\em this} impression that we will rebut. We of course admit that in general, (Local) gauge-dependent structures may be `invisible from the other side'. But surprisingly, for the case of gauge/gravity duality: some (Local) gauge structure {\em is} visible, and our analysis characterises the class of visible diffeomorphisms.\footnote{Diffeomorphisms that are (Invariant) but not (Fixed), see Section \ref{noinvis}.} This is our result in Section \ref{invarces}: roughly, that each of a certain class of (Local) gauge symmetries of the gravity/bulk theory, viz.~diffeomorphisms,  is mapped by the duality to (Redundant) gauge symmetries of its dual, viz.~to position-dependent conformal symmetries of the conformal field theory defined on the boundary.

There are two reasons why this result is worth stressing. First, we think the false impression is widespread (presumably because of the above line of thought, that preserving the gauge structures `would be even more striking'). In particular, we shall see that no lesser authors than Horowitz and Polchinski seem to endorse it.\footnote{In Horowitz and Polchinski~(2006), it is left ambiguous whether the putatively invisible symmetries should be interpreted as only (i)  (Local) or as also (ii) having the additional boundary condition of ``asymptotic triviality''. 
The exact interpretation will not matter for our argument, since we will precisely characterise, in Section \ref{noinvis}, which symmetries are visible, and which are invisible, and the conditions that they satisfy.}

Second, we will see at the end of Section 6 that there is an interesting connection (and contrast) with Einstein's infamous hole argument. Roughly: whereas Einstein's hole argument turns on there being a plethora of diffeomorphisms of a certain required type, our `bulk argument' turns on the required type of diffeomorphism being restricted to the invisible ones. Thus recall that Einstein uses an arbitrary diffeomorphism in a certain region of spacetime,`the hole', which is chosen (so as to make the argument a rebuttal of determinism) to lie to the future of some time-slice. So his diffeomorphisms are required to be the identity on the time-slice (and to its past). Our result also has a preferred surface of codimension one (albeit not a spacelike surface, but rather, a timelike cylinder): the boundary, at which a diffeomorphism is required to preserve the fields, so as to be `invisible' to the theory defined there.

 
\section{Duality as a symmetry between theories}\label{thydlty}

\subsection{Motivating the definitions}\label{motivating}

In this Section, we propose schematic definitions of a physical theory, and of a duality. The definitions will be general enough to apply to both classical and quantum physics, though we will of course have quantum physics, especially quantum field theory  and string theory,  mostly in mind. The definitions will also be simple, you might even say simplistic. For example: (i) we will take a theory  to have a single state-space, and so to deal with a single type of system; (ii) we will not mention coupling constants, and so we will not require a duality to relate weak and strong couplings. But of course discussion tends to be clearer with simple definitions: which can then be qualified or added to, as needed. And we will see in Section \ref{illustrates} that our simple definition of duality is indeed instantiated by our main case, gauge/gravity duality.\footnote{More precisely, we should say, so as to allow for the duality's path integrals etc.~not being (yet!) rigorously defined: our definition is instantiated {\em formally}.}    

The idea of the definitions is that duality is a symmetry between theories. But while symmetry is a matter of sameness between distinct situations, according to a theory, (e.g.~sameness of quantities' values on distinct states, for an active symmetry): duality is a matter of sameness between theories. 

Of course, we will need to be careful about when theories are the same (or `theoretically equivalent'); and we will need to allow that formal isomorphisms do not in general imply sameness of content, `saying the same thing'. Besides, we will need to respect three closely related points about `sameness' for symmetries.\\
\indent  (1): We will need to respect the contrast between (i) an active symmetry, i.e.~a map sending each state to another that has the same values for each quantity, and (ii) a passive symmetry, i.e.~a map sending each quantity to another that assigns the same value to each given state.\\
\indent  (2): We should bear in mind that a symmetry, whether active or passive, usually concerns a {\em subset} of quantities, since the set of all quantities usually separates the states (in the sense that any two states differ in their values of at least one quantity). Thus under an active symmetry such as spatial translation, most quantities' values are preserved---but the values of position quantities are of course changed; and similarly, under a passive symmetry, usually not all quantities get mapped.\\
\indent  (3): We should bear in mind the idea of gauge, especially in Section \ref{samedifferent}'s general philosophical meaning (Redundant), and associated ideas like gauge symmetry. For example, building on (1) and (2), we can say: an active gauge symmetry, i.e.~a map that leaves the gauge-equivalence classes (gauge-orbits) invariant, implies that    the gauge-invariant quantities do not separate the states.

But in this paper, we can for the most part take these aspects in our stride---even the last, (3), about gauge. For to make our two main points, we will not need to articulate them formally. 

So a duality is, roughly, an `isomorphism' between two theories, or from a theory to itself. This `isomorphism' will act in a `meshing' way on the theories' (or theory's)  state-spaces, collections of quantities, and dynamics.

Besides, to develop (1)'s idea of a passive symmetry, one envisages that some special subset of quantities represents `descriptive apparatus'. For example, the subset could  be given by  a set of charts on a differential manifold (e.g.~for a classical mechanical state-space), or  by a set of bases in a Hilbert space. Then we are able to distinguish cases where: (a) the duality changes the physical state, and so the values of (some) quantities, so that it is `active'; as against (b) the duality changes only descriptive apparatus, and so is `passive'. And besides, we are able to talk about gauge as descriptive redundancy.  But as mentioned: in this paper, we can for the most part take the active/passive contrast in our stride, and so not articulate  `descriptive apparatus' formally.    

To make these ideas more precise, we define: first theories (Section \ref{defth}), and then duality maps (Section \ref{duality}). 

\subsection{The definition of `theory'}\label{defth}

We take a {\em theory} as a triple $T = \langle {{\cal S}}, {{\cal Q}}, D \rangle$ of state-space $\cal S$, set of quantities $\cal Q$ and dynamics $D$. States and quantities are assignments of values to each other. So there is a natural pairing, and we write $\langle Q ; s \rangle$ for the value of $Q$ in $s$. In classical physics, we think of this as the system's intrinsic possessed value for $Q$, when in $s$; in quantum physics, we think of it as the (orthodox, Born-rule) expectation value of $Q$, for the system in $s$.\footnote{This idea of an quantity-state pairing, and associated ideas about quantities separating states, and {\em vice versa},  is of course  widely used (and often called `duality'!) in broadly structural studies of classical  and quantum physics.} When we use this formal structure in Section \ref{ggd}, we will denote the pairings $\bra Q;s\ket$ as $\bra s|Q|s\ket$. 

As to the dynamics, $D$: this can be taken as the time-evolution of states (and so quantities' values), or as the time-evolution of quantities (and so values) for a fixed state. In quantum physics, the alternative pictures are of course  called the `Schr\"odinger-picture' and the `Heisenberg-picture'; but the pictures apply equally to classical physics. We will adopt the more familiar Schr\"odinger-picture; and also take the dynamics to be deterministic. So time-evolution is given by a group action $D^{\tn{Sch}}$ of $\mathR$ on ${\cal S}$, with $s$ the state at fiducial time $t=0$, $ \mathR \times {{\cal S}} \ni (t,s) \mapsto  D^{\tn{Sch}}(t,s) =: s(t) \in {{\cal S}}$.

We should comment on our taking a theory  to have a single state-space, and so to deal with a single type of system. There are three main points to make: each will lead in to the next.

(1): {\em A `via media'}: Our notion of theory is: (i) more general than one prevalent understanding of `model of a theory'; but also, (ii) more specific than one prevalent understanding of `theory'. Let us  spell out this {\em via media} construal of `theory'.\\
\indent \indent  (i): Our notion is more general than a specific history of a system. Such a history corresponds (on the Schr\"odinger-picture) to a sequence of instantaneous states, i.e.~a curve through the state-space; and is often called a `model' of the theory.  \\
\indent \indent  (ii): On the other hand, our notion is more specific than a set of such state-spaces, each equipped with quantities and dynamics. But such a set is a prevalent understanding of `theory', both in physics and in philosophy. In physics, for example: the theory of Hamiltonian mechanics is often taken as the set of all appropriate state-spaces (e.g.~all finite-dimensional cotangent bundles), with their associated quantities (e.g.~the measurable real functions on the bundle); and similarly for many other theories, e.g.~Lagrangian mechanics as the set of tangent bundles, quantum mechanics as the set of separable Hilbert spaces with their bounded operators. And in philosophy: the semantic conception of scientific theories takes a theory to be a set of models (of course, often less well-defined than these physics examples).\footnote{\label{Halv}{For how to conceive the relation between Hamiltonian and Lagrangian mechanics, understood to include all the appropriate state-spaces, as itself an isomorphism, cf.~Teh and Tsementzis (2015). As to the semantic conception of theories, Halvorson (2015) is a recent discussion: one of his main morals, which we endorse, is that the debate between the syntactic and semantic conceptions is in various ways misdirected. For the connection to our topic of theoretical equivalence cf.~also Barrett and Halvorson (2015) and references therein.}} 

(2): {\em `Distinct but isomorphic'}: The more general notion of theory, in (1) (ii) above, prompts yet another comment on the variety of meanings of `saying the same thing'. For that notion's generality makes vivid that a theory usually treats many systems, including systems of different  types that require very different  state-spaces. The obvious examples are that the state-spaces for classical Hamiltonian, or for Lagrangian, or for quantum, systems can differ in their dimension. So we stress that while our {\em via media} construal of `theory' restricts a theory to a specific state-space:\\
\indent (a): we accept that this is against a common usage of `theory'; but it will nevertheless be convenient for us; \\
\indent (b): our usage allows us to recognize---and we of course do recognize---that usually there are many {\em token} systems of the type treated by a theory in our sense (i.e.~the type  specified by a given state-space, set of quantities and dynamics). 

Point (b) prompts an important general point.  We of course recognize (as everyone must!) that there are cases of two disjoint parts of reality---in Hume's phrase, two `distinct existences'---that: match exactly,  `are isomorphic', in the taxonomy used by some theory (i.e.~as regards the properties and relations the theory treats)---but are otherwise different, i.e.~distinct and known to be distinct. This point has an obvious corollary for what in Section \ref{intro} we announced as this paper's first main point: that dual theories `can say the same thing'. For it is natural to take `saying the same thing' to mean saying (i) the same assertions about (ii) the same  objects. Then we must beware that a definition of dual theories as `isomorphic' (as in Section \ref{duality} below) will cue in only to (i); and the possibility of `distinct but isomorphic existences' implies that isomorphism alone does {\em not} secure (ii). 

So we stress that when we declare dual theories `can say the same thing' (and duality is thereby like gauge symmetry), we do {\em not} mean to ignore this corollary---to ignore distinct but isomorphic existences, and to assume that (ii) is secured. We acknowledge that (ii)---the theories' being about the very same systems---is a matter that outstrips the essentially formal relation given by the duality map. Hence our use of the word `can'! This leads in to the next point.\footnote{These remarks also bear on criteria for theoretical equivalence that cue in only to aspect (i) above, like those discussed in the references cited in footnote \ref{Halv}. Such formal criteria can provide at best necessary conditions of `saying the same thing'. For they do not secure (ii), that the theories are about the very same objects.}   

(3) {\em `Theories of the universe'}: But there is one kind of scientific context where the idea of distinct but isomorphic existences falls by the wayside: namely, when we aim to write down a cosmology, i.e.~a theory of the whole universe. For such a theory, there will be, {\em ex hypothesi}, only one token of its type of system, viz.~the universe. (Besides, the theory will in principle aim to state enough about the relations between the many sub-systems of the universe that no two sub-systems will get exactly the same description; so that even at the level of sub-systems, there will not be distinct but isomorphic existences.) Agreed, this scientific context is very special, and very ambitious: we rarely aim to write down a cosmology. But of course, it {\em is} the context of much work in string theory, and quantum gravity more generally. So it will apply when we turn, in Section \ref{ggd}, to string theory, and in particular to gauge/gravity duality.\footnote{In this context, one might go further than setting aside the possibility of distinct but isomorphic existences. One might also hold that the interpretation of our words, i.e.~of the symbols in the cosmological theory, must be fixed from within the theory: this view is endorsed, under the label `internal point of view', in De Haro (2015, Section 2.4.1) and Dieks et al. (2015, Section 3.3.2). We should note however that a lot of work on gauge/gravity duality concerns systems that are {\em much} smaller than the universe. For gauge/gravity ideas turn out to be very useful for understanding strongly coupled microscopic systems, like a quark-gluon plasma as explored by instruments like the Relativistic Heavy Ion Collider. Recall Section 2's point (2) about dualities being useful; and cf.~e.g.~McGreevy (2010).} 

\subsection{The definition of `duality'}\label{duality}

So much by way of discussing our notion of `theory' and its consequences. To motivate, and to aid comparison with, our definition of duality, we now briefly discuss the idea of a symmetry of a theory. We may define a {\em symmetry} as a map $a:{{\cal S}} \raw {{\cal S}}$ (we write `$a$' for `automorphism') that preserves the value of the quantities, i.e.
\be
\langle Q ; a(s) \rangle = \langle Q ; s \rangle \; .
\label{symmypresvalue0sec2}
\ee
As mentioned in items (1) and (2) of Section \ref{motivating}, this notion of symmetry: (i) is `active', a map on states, and (ii) would usually preserve the values of only a certain subset of the quantities.  Item (1)'s idea of a passive symmetry can then be defined using the idea of a {\em dual map}. Thus, given any map $a:{{\cal S}} \raw {{\cal S}}$ (not necessarily an active symmetry), the dual map on quantities, $a^*:{{\cal Q}} \raw {{\cal Q}}$, is defined by $\langle a^*(Q) ; s \rangle := \langle Q ; a(s) \rangle$. Then, assuming that $a$ is a symmetry  as in \ref{symmypresvalue0sec2}: we can define the passive symmetry as just the dual map $a^*$  on quantities, by 
\be
\langle a^*(Q) ; s \rangle := \langle Q ; s \rangle \equiv \langle Q ; a(s) \rangle \; .
\label{symmyqtydualdef0}
\ee\\


We now define a duality as a `meshing' map between two theories. $T_1 = \langle {{\cal S}_1}, {{\cal Q}_1}, D_1 \rangle$ is {\em dual} to $T_2 = \langle {{\cal S}_2}, {{\cal Q}_2}, D_2 \rangle$ iff there are bijections $d_s: {{\cal S}_1} \raw {{\cal S}_2}, d_q: {{\cal Q}_1} \raw {{\cal Q}_2}$, ($d$ for `duality'), such that: \\
 (i) they give matching values of quantities on states in the sense:\footnote{In Section \ref{illustrates} we will impose a stronger condition, viz.~the values of the quantities on any {\it pair} of states match: $\bra Q_1;s_1,s_2\ket_1=\bra d_q(Q_1);d_s(s_1),d_s(s_2)\ket_2$, $\forall s_1,s_2\in{\cal S}_1$. This strengthening of our simple approach in this section is natural for quantum theories, because it amounts to unitary equivalence.}
\be\label{obv1}
\langle Q_1; s_1 \rangle_1 = \langle d_q(Q_1) ; d_s(s_1) \rangle_2 \; , \;\; \forall Q_1 \in {{\cal Q}_1}, s_1 \in {{\cal S}_1}; 
\ee
 (ii) $d_s$ commutes with (is equivariant for) the two theories' dynamics; i.e.~in Schr\"odinger-picture, the group actions $D^{{\tn{Sch}}_i}$ of $\mathR$ on ${\cal S}_i$:
\be\label{obv2}
d_s(s_1(t)) \equiv  d_s(D^{{\tn{Sch}}_1}(t,s_1)) = D^{{\tn{Sch}}_2}(t,d_s(s_1)) \; , \;\; \forall t \in \mathR, s_1 \in {{\cal S}_1}. 
\ee
Eq. \ref{obv1} and \ref{obv2}  appear to be asymmetric between $T_1$ and $T_2$. But in fact they are not, thanks to the maps $d$ being bijections. 

Thus we have adopted the definition of `duality' that is obvious and simple, given our conception of `theory'. One  could strengthen the definition in various ways: for example, to require that $d_s$ be a symplectomorphism for Hamiltonian theories, unitary for quantum theories etc.  And there is a whole tradition of results relating the requirement of matching values \ref{obv1} to such strengthenings (the obvious one in quantum physics being Wigner's theorem that the map's preserving all the transition probabilities implies its being  unitary or anti-unitary). But we do not need to pursue such strengthenings. 

It is not just that, as we said at the start of Section \ref{motivating}: a simple definition is clearer, and can be weakened and qualified, as needed. Also, with this definition, we immediately establish our first main point: that two dual theories can be gauge related, in Section 2's general philosophical sense, (Redundant), of being physically equivalent. More precisely: the point follows `immediately', when we bear in mind our preceding discussion, especially the emphasis in Section \ref{defth} points (2) and (3),  on the {\em can be} in `can be gauge related': i.e.~our allowance of distinct but isomorphic `existences', and how this allowance falls by the wayside for a cosmology, i.e.~a theory of the whole universe.

And more important: we will see in Section \ref{ggd} that this simple definition is indeed instantiated, albeit formally, by gauge/gravity duality---and using maps $d_s, d_Q$ that are (formally) unitary. To this topic, we now turn ...  \\
   
\section{Gauge/gravity duality}\label{ggd}

We will first (Section \ref{IntroAdSCFT}) give a brief introduction to the original and most-studied case of gauge/gravity duality: AdS/CFT, the duality between a gravity theory on anti-de Sitter spacetime (`AdS') and a conformal field theory (`CFT') on its boundary. Then in Section \ref{illustrates}, we will show how to calculate quantities on both sides, and argue that Section \ref{duality}'s simple definition of duality is indeed instantiated, albeit formally, by AdS/CFT.

\subsection{Introducing AdS/CFT}\label{IntroAdSCFT}

\subsubsection{The two sides}\label{2sides}

There are many pedagogic introductions to AdS/CFT: for example, Maldacena's own pedagogy (2004, 2004a) is excellent.  In this subsection, we will follow Section 2 of McGreevy (2010), which follows the line of thought of the first half of Horowitz and Polchinski (2006). Cf.~also e.g.~De Haro et al.~(2016). There is also an excellent book, Ammon et al.~(2015). 

The general idea of gauge/gravity duality is that some gauge quantum field theories (`QFTs') in $d$ spacetime dimensions  are dual to some quantum theories of gravity in a $(d+1)$-dimensional spacetime that has the $d$-dimensional manifold of the QFT as its conformal boundary.
Hence the jargon: gauge on the boundary; gravity in the bulk. Besides: in some examples, the space-time is not anti-de Sitter, and the quantum field theory is not conformal (or even relativistic!). So all hands agree that `gravity/gauge' would be a better overall name than `AdS/CFT'.

The details of the `dictionary' (equivalence, or correspondence, of theories) of course vary from one case (pair of theories) to another. But McGreevy, Horowitz and Polchinski describe three hints that apply across the cases.\\
\indent (1): A no-go theorem of Weinberg and Witten states, roughly speaking, that a Poincar\'e-invariant QFT cannot contain a graviton (a massless spin-2 particle). With the benefit of hindsight, we can say that this theorem suggests that to `get' a graviton within that framework, we should increase the number of spacetime dimensions we consider.\\
\indent (2): Black hole thermodynamics suggests that the maximum entropy in a region of space is given by the area of its boundary, in Planck units; (this maximum being attained by a black hole). This is much smaller than the entropy we would expect to be associated to a local QFT defined on the region (even if it has some UV cutoff): which we would expect to scale with the volume. This hints that a quantum theory of gravity should have a number of degrees of freedom which scales like the number of degrees of freedom in a QFT in one lower spatial dimension.\\
\indent (3): In a QFT, the beta-function (which describes how the couplings $g(u)$ depend on the energy scale $u$, viz.~by the equation $u\, \partial_u g = \beta\left(g(u)\right)$) is local in $u$. This can be traced to spacetime locality (i.e.~local couplings in the Lagrangian). But it hints that $u$ itself be taken as the extra spatial dimension, whose addition to the framework of the QFT defines the bulk.

To exploit these hints to try and find the simplest concrete example, let us assume the beta-function of our QFT in $d$ dimensions vanishes, i.e.~the system is self-similar: so the scale transformation $x^{\mu} \mapsto \lambda\, x^{\mu}, (\mu = 0, 1, 2,..., d-1)$, is
a symmetry. If we now propose, following the hints, to add an extra dimension $u$ which is an energy scale, then $u$ should transform under this dilation as $u \mapsto u/{\lambda}$. And it can be shown that the most general ($d+1$)-dimensional Poincar\'e-invariant metric with this symmetry is the anti-deSitter metric, written AdS$_{d+1}$. This metric can be put in the following form, where $\ell$, the `AdS radius', is 
a length, and $r := \ell^2/u$:
\be
\dd s^2 = {\ell^2\over r^2} \left(\dd r^2+ \eta_{\mu \nu}\,\dd x^{\mu}dx^{\nu}\right) .
\label{AdSmetric}
\ee
Pictorially, AdS$_{d+1}$ is a family of copies of $d$-dimensional Minkowski spacetime of varying size. The family is parameterized by $r$. The boundary is the locus $r \raw 0$, corresponding to the UV sector of the QFT.

The overall situation is now that the metric \ref{AdSmetric} solves the equations of motion of an action of the form
\be
S_{\tn{bulk}}[g,...] = \frac{1}{16 \pi \GN^{(d+1)}} \, \int \dd^{d+1}x \, {\sqrt g} \left(R-2 \Lambda + \ldots\right) \; ,
\label{SB}
\ee
and that different QFTs on the boundary $d$-dimensional Minkowski spacetime correspond to different theories of gravity in the bulk, varying in their field content and bulk actions---and the main role of string theory is to provide consistent ways of filling in the dots in \eq{SB}. More details in Section \ref{dicty}. But we can already say that in \eq{SB}:\\
 \indent (i) $R$ is the Ricci scalar (so that ${\sqrt g} \,R$ is the Einstein-Hilbert action for classical gravity), \\
\indent (ii) $\GN^{(d+1)}$ is the $(d+1)$-dimensional Newton's constant, \\
 \indent (iii) the cosmological constant $\Lambda$ is related to the AdS radius by $-2 \Lambda = d(d-1)/\ell^2$,\\
 \indent (iv) the dots represent various higher powers of curvature, boundary terms, and-or other bulk fields.

Also, we can already write down an entry common to the `dictionaries' between two such theories, one in the bulk and one on the boundary. Namely, the correspondence between the radial coordinate $r$ in the bulk, and the RG flow in the boundary (with $r = 0$ corresponding to the UV sector): \\
\\
\indent \indent \indent radial motion in coordinate $r$ in AdS $\leftrightarrow$ RG flow in QFT.\\
\\
Although in this exposition, this correspondence is in effect postulated by hint (3) above, it is {\em a priori} very remarkable.

The original case (Maldacena (1998): which is also nowadays the best studied case\footnote{Previous work gave (at least with the benefit of hindsight!) various intimations or hints  of this case. And Maldacena (2004, p. 66; 2004a, p. 161) and Horowitz \& Polchinski (2006, p. 6) duly note that this case of the duality was presaged by earlier work by 't Hooft and by Polyakov. In particular, 't Hooft thought that interacting gluons with a large number of colours would give rise to a string theory, with the strings composed of interacting gluons. The work of Brown and Henneaux (1986), relating the symmetries of gravity in three-dimensional asymptotically AdS to the Virasoro algebra of a two-dimensional CFT, is another early hint of the duality: cf.~Section \ref{gauge}.}) takes the QFT to be a strong coupling regime of a supersymmetric  gauge theory (`SYM' for `super Yang-Mills') in four spacetime dimensions ($d$ = 4), with gauge group $\mbox{SU}(N)$ with large $N$. Here:\\
 \indent (1): $N$ is the number of colours (sometimes written $N_c$);\\
 \indent (2):  the theory uses four copies of the minimal supersymmetry algebra (written ${\cal N} = 4$);\\
 \indent (3): `strong coupling' means that the product $g^2_{\tn{YM}}N$ (called the `'t Hooft parameter'; $g_{\tn{YM}}$ is the theory's dimensionless coupling) is large;  and\\
 \indent (4): `large $N$' means that we take the limit $N \raw \infty$ and $g_{\tn{YM}} \raw 0$ with $g^2_{\tn{YM}}N$ fixed. (This is called the `'t Hooft limit', and an expansion in the parameter $1/N$ is called a `'t Hooft expansion'; Butterfield and Bouatta (2015) is a philosophers' introduction.)\\
This theory has the special feature that the beta-function vanishes, and the theory is conformally invariant.  

This theory's dual, on the bulk side, is: a supersymmetric gravity theory on AdS$_5$, which arises by compactification of 10-dimensional IIB string theory on AdS$_5 \times S^5$. (Here `compactification' means that, coordinatizing the ten dimensions as $x^0, x^1, ... , x^9$, fields are classified according to their harmonic dependence on the coordinates, $x^5, ... , x^9$, which parametrise the $S^5$: and at low energies, we keep the leading contribution, from the lowest-lying harmonics: the modes with no dependence on the $S^5$ coordinates.) This supersymmetric gravity theory takes the form \eq{SB} and depends on two parameters:\\
\indent (a): the five-dimensional Newton constant $\GN^{(5)}$: which is the 10-dimensional Newton's constant\footnote{The 10-dimensional Newton's constant is given by the parameters of the string, as: $\GN^{(10)}=8\pi^6g_s^2{\a'}^4$, where $g_s$ is the string coupling (cf.~Section \ref{dicty}, after \eq{QEA}) and $\a' :={1\over2\pi T}$, where $T$ is the tension of a fundamental string. $\a'$ therefore has units of a length squared.\label{parameters}} $\GN^{(10)}$, divided by the volume of the internal sphere; and \\
\indent (b): the radius of the $S^5$, $\ell$, which also determines the curvature radius of the AdS space (through $\Lambda=-6/\ell^2$: cf.~(iii) above).

\subsubsection{The dictionary}\label{dicty}

To set up the dictionary, let us first see how the duality relates the {\em free parameters} on the two sides.

 On the bulk side there are two free dimensionless parameters: the string coupling $g_s$ and the ratio of the two dimensionful parameters $\ell$ and $\GN^{(5)}$, mentioned at the end of Section \ref{2sides}. Both $\GN^{(5)}$ and $\ell$ are measured in units of the squared string length  ${\alpha}'$ (cf.~footnote \ref{parameters}). On the SYM side there are also two dimensionless parameters: $N$ and $g_{\tn{YM}}$ (cf.~points (1)-(4) above). The relation between the two sets of constants is: (a) $g_s=g_{\tn{YM}}^2/4\pi$, (b) ${\GN^{(5)}/\ell^3}={\pi\over2N^2}$, (c) $\ell^4/{\alpha'}^2=g_{\tn{YM}}^2N$. Thus (cf.~point (3) above), strong coupling on the SYM side corresponds to a large value of $\ell^4/\alpha'^2$, and weak coupling corresponds to a small value. Recall that $\ell$ is both the $S^5$ radius and the AdS$_5$ radius of curvature. Therefore, the regime of strong SYM coupling corresponds to small bulk curvatures (the regime in which general relativity is a good approximation) whereas weak SYM coupling corresponds to large bulk curvatures (and in that case general relativity cannot be trusted---compare what happens near the singularity of a black hole). This instantiates our earlier characterisation (2), in Section \ref{samedifferent}, of the duality as being `useful'.  

The {\em symmetries} also agree. The symmetry group of AdS$_5\times S^5$ is $\mbox{SO}(2,4)\times\mbox{SO}(6)$. The first factor is indeed the conformal group in four dimensions; and the Lie algebra of SO(6) is isomorphic to that of SU(4), which corresponds to the rotations of the six scalars and four fermions in the SYM (recall, cf.~(2) at the end of Section \ref{2sides}, that there were four copies of the minimal supersymmetry algebra). So the symmetries also match. In a moment, we will see how the quantities match.

Before we do that, let us comment on the {\em quantum corrections}. These appear as higher-order corrections in the action \eq{SB}, which is regarded as a low-energy effective action. We write the first of these for the 10-dimensional type IIB supergravity action:
\bea\label{QEA}
S={1\over16\pi \GN^{(10)}}\int\dd^{10}x\,\sqrt{-g}\left(R+\gamma(\phi)\,W+\ldots\right)~,
\eea
where the dots denote the matter fields. $W$ here is a scalar obtained by contracting four powers of the Weyl tensor; it thus contains eight derivatives, whereas the leading term contains two. The coefficient $\gamma(\phi)$ is thus of length dimension six: indeed, it is proportional to ${\alpha'}^3$. $\phi$ is the dilaton field, whose asymptotic value gives the coupling on the string world sheet $g_s$ mentioned earlier. 

The above quantum effective action contains two kinds of corrections:\\
(a) Higher-derivative corrections: higher-curvature corrections, of third order in $\alpha'$. These are genuine perturbative quantum gravity corrections (where $\alpha'$ plays the role of $\hbar$). We should thus regard this action as the quantum effective action, to the given order, of type IIB string theory. The parameter of the quantum expansion is $\alpha'^3$, which after compactification gets multiplied by factors of $1/\ell^6$. Thus, the order of these corrections to the leading AdS$_5$ results is $1/N^{3/2}$. This is indeed a small number in the 't Hooft limit.
\\
(b) String theory corrections in $g_s$. These arise from the string theory perturbative expansion (an expansion in world-sheet topologies). In fact, in type IIB, the entire series is known, including the non-perturbative contributions: due to the $\mbox{SL}(2,\mathbb{R})$ symmetry enjoyed by type IIB supergravity.  In string perturbation theory the $\mbox{SL}(2, \mathbb{R})$ symmetry is broken by the vacuum expectation value of the dilaton, but it is believed that the quantum theory possesses a local $\mbox{SL}(2,\mathbb{Z})$ symmetry. Because the quantum effective action \eq{QEA} possesses this {\it discrete} symmetry, the symmetry must be preserved order by order in $\alpha'$ (Green (1999)). This in fact allows one to determine the function $\gamma(\phi)$ as a particular modular form.\footnote{The Eisenstein series $E_{3/2}$.} This can then be compared to the quantum $\mbox{SL}(2,\mathbb{Z})$ symmetry (Montonen-Olive electric-magnetic duality) of the SYM theory. This is one of the non-trivial (and non-perturbative) tests of AdS/CFT. \\

We have so far discussed the original case, of the type IIB/SYM correspondence, and some of the checks of the correspondence based on parameters and symmetries. Let us now discuss some cases that exhibit the correspondence between {\em quantities}, and thereby the dynamics, of the two theories. We generalise to arbitrary boundary dimensions $d$. 

It turns out that there is a relatively simple formulation of the AdS/CFT dictionary,\footnote{From now on, except where we indicate it explicitly, we will be dealing with the Euclidean version of AdS/CFT, which is the most studied in the literature. That is, both the asymptotic bulk geometry and the CFT background are spaces of Euclidean signature.} which can be written down easily for a scalar field, and which also generalises to tensor fields. Take the path integral for the scalar field $\phi$, in Euclidean signature, as a function of the boundary conditions:\footnote{For simplicity, we are now taking the metric to be fixed and we are suppressing it in the notation.}
\be\label{bulk}
Z_{\tn{string}}[\phi_{(0)}] := \int_{\phi(r,x)|_{\tn{bdy}}=\phi_{(0)}(x)}{\cal D}\phi~\exp\left(-S_{\tn{bulk}}[\phi]\right) \; .
\ee
The AdS/CFT correspondence now declares that this equals the generating functional of correlation functions in the dual CFT:
\bea\label{adscft}
Z_{\tn{string}}[\phi_{(0)}]\equiv \Big\bra\exp
\left(\int\dd^dx~\phi_{(0)}(x)\,{\cal O}(x)\right)\Big\ket=:Z_{\tn{CFT}}[\phi_{(0)}]~,
\eea
where the expectation value  is evaluated in the vacuum state of the theory. 

Part of the above recipe is the field-operator correspondence: for each bulk field $\f(r,x)$ (where $\f$, and thus \eq{bulk}-\eq{adscft}, are now interpreted to hold for {\it any} bulk field) of a given mass and spin, there is a corresponding gauge-invariant operator ${\cal O}(x)$ of a certain scaling dimension and spin (in a CFT, the dimension uniquely determines a gauge-invariant, scalar operator). In particular, the asymptotic value of the field, $\phi_{(0)}(x)$, is a `source' that couples to ${\cal O}(x)$. 

For $p$-form fields (for scalars $p=0$, for vectors $p=1$), the scaling dimension $\Delta$ is related to the mass of the bulk scalar field and the boundary dimension by:
\bea\label{dim}
\D={d\over2}+\sqrt{{(d-2p)^2\over4}+m^2\ell^2}~.
\eea

In the next two subsections we will show how to calculate quantities (correlation functions) on both sides of the duality, using the two sides of \eq{adscft} independently, thus checking the announced correspondence. We now give two examples that illustrate: (i) the field-operator correspondence that appears in \eq{adscft}, subject to \eq{dim}; (ii) how gauge invariance is treated independently on both sides of the correspondence (we will take up this topic again in Section \ref{GIduality}). 

The first example of the correspondence between bulk fields and boundary operators in \eq{adscft} is a specific gauge-invariant operator with dimension $\Delta=4$, constructed for $d=4$ from the Yang-Mills field strength $F$: ${\cal O}(x)={1\over4g_{\tn{YM}}^2}\,\mbox{Tr}\,F^2(x)$. This basic operator corresponds, through \eq{dim} ($d=4$, $p=0$, $m=0$), to a massless scalar field in AdS$_5$. Thus, the correspondence relates scalar fields to gauge-theoretic operators, but in a gauge-invariant way.

In our second example, we consider the Maxwell action for a U(1) gauge field in AdS$_4$, rather than a scalar field. From \eq{dim} with  $d=3$, $p=1$, $m=0$, we get $\D=2$. The corresponding CFT operator is a conserved vector current $J_\m$, which is  indeed of dimension 2. $J_\m$ couples to a source $A_\m$ for a global symmetry in the exponential in \eq{adscft} in the CFT. It is straightforward to check, from \eq{adscft}, that the one-point function $\bra J_\m\ket$ equals the bulk electric field (again, a gauge-invariant quantity corresponding to the U(1) gauge field). 

\subsection{AdS/CFT exemplifies the definition of duality}\label{illustrates}

We now turn to  how calculate quantities from \eq{adscft} and how this formula instantiates our  definition of duality, viz.~\eq{obv1} and \eq{obv2}. We begin with the boundary, CFT, side (Section \ref{bdy...}). Then we describe how the bulk side matches it (Section \ref{matches}). This will also clarify the meaning of \eq{bulk}; and this clarification will furnish a reply to an objection (Section \ref{obj}).

\subsubsection{The boundary side ...}\label{bdy...}

In a CFT, the quantities one is interested in  are the correlation functions:
\bea\label{correlators}
\bra s'|\,{\cal O}(x_1)\cdots{\cal O}(x_n)\,|s\ket~,
\eea
expectations of products of gauge-invariant operators with dimension $\Delta$, for some states $s, s'$. For instance, to calculate the two-point function in the state $s=s'=0$, i.e. the expectation of a binary product in the vacuum state, one takes the functional derivative of \eq{adscft} twice and sets the source to zero. The result can be shown to be:
\bea\label{2pt}
\bra{\cal O}(x)\,{\cal O}(y)\ket_{\rm{vac}}= Z[0]^{-1}\,{\delta^2Z[\phi_{(0)}]\over\delta\phi_{(0)}(x)\,\delta\phi_{(0)}(y)}\Big|_{\phi_{(0)}=0}={1\over|x-y|^{2\Delta}}~,
\eea
where $|x-y|$ is the distance between the two boundary points. (In the next subsection, we will also calculate this correlation function using the bulk theory.)

This can be generalised to calculate general vacuum expectation values \eq{correlators}: one takes functional derivatives of the right-hand-side of \eq{adscft} with respect to the source, and sets the source to zero at the end of the calculation. Expectation values in a general state $|s\ket$ can be evaluated using the state-operator correspondence that holds in any CFT: the states are classified by the representations of the conformal group, and the latter can be obtained by applying the unique gauge invariant operator, corresponding to it, to the vacuum.\footnote{For simplicity, we have limited ourselves here to scalar operators and the corresponding states, which are eigenstates of the dilatation operator. This readily generalizes to operators with non-zero spin.} Thus  the gauge theory side of the duality is automatically stated in the language of states and quantities that was used in Section \ref{defth} to define a theory.

Some cautionary remarks should be added here. For a generic QFT, a formula such as the right-hand side of \eq{adscft} is  formal: to define it rigorously requires knowledge of the full non-perturbative structure of the theory, which obviously is available only for very few quantum field theories (mainly: topological QFT's, and field theories in two dimensions). 

But for the field theories that are involved in the basic examples of AdS/CFT, viz.~gauge theories with vanishing beta function (e.g.~the original case of ${\cal N}=4$ SYM discussed in \ref{IntroAdSCFT}),\footnote{In the case of deformations away from conformality, through the addition of operators that break conformal invariance (which correspond to a modification of the boundary conditions in the bulk theory), conformal symmetry in point (ii) below only holds near a fixed point of the RG flow. See e.g.~Witten (2001), pp. 8-10.} the situation is slightly better than for generic QFTs, for two reasons:\\
\indent (i): There are perturbative methods, such as the 't Hooft expansion (cf.~Section \ref{IntroAdSCFT}), that allow a fully quantum exploration of the Hilbert space. In this approximation, the quantum corrections are included as a systematic series in $1/N$. On the string theory side (and in a different regime of 't Hooft coupling), these corrections appear as a series of higher-order curvature corrections in the supergravity action, as we saw in \eq{QEA}, and discussed there under point (a).\\
\indent (ii): Conformal symmetry and supersymmetry strongly constrain the operators and the states, so that some non-perturbative results (such as conformal anomalies, non-renormalization theorems, and instanton contributions to certain amplitudes) are known. And in general, it is expected that the above formal structures will some day be defined with rigorous mathematics. 

Let us give two examples that illustrate both points (i) and (ii). The first has already been mentioned,  in Section \ref{dicty} under (b): the exact $\mbox{SL}(2,\mathbb{Z})$ symmetry enjoyed by both type IIB and SYM theory. This allows one to calculate the coefficient $\gamma(\phi)$ defined in \eq{QEA}, non-perturbatively in the string coupling. 

Another such result is the conformal anomaly, which we quote here for $d=4$ but which is known in all dimensions:
\bea\label{anomaly}
\bra T^i{}_i\ket={N^2\over32\pi^2}\left(R^{ij}R_{ij}-{1\over3}\,R^2\right)~.
\eea
In a CFT, the trace of the stress-energy tensor is zero, classically. For an even number of dimensions $d$, the vacuum expectation value of the stress-energy tensor acquires a non-zero value \eq{anomaly}, which is not suppressed, but indeed {\it enhanced} in the large-$N$ limit---notice the $N^2$-dependence! This anomaly is an exact result: it does not rely on any perturbative scheme, since it is in essence a geometric result (Deser et al.~(1993)), and the contributions that appear on the r.h.s.~in various dimensions have been classified in the literature as topological and geometric invariants. 

\subsubsection{... Matches the bulk side}\label{matches}

In order to substantiate our claim---that Section \ref{duality}'s simple definition of duality is indeed instantiated by AdS/CFT---we still need to show that the bulk side of the story, as in \eq{bulk}, can also be written in the language of states and operators. Obviously, the same warnings issued after \eq{2pt} apply here: in the absence of a rigorous definition of the path integral, or of a non-perturbative definition of the Hilbert space, our approach is obviously heuristic and, strictly speaking, valid only in a perturbative expansion in $1/N$. (But there are a few non-perturbative results, which we will mention below: and the expectation is that the formalism of states and operators, with appropriate modifications, still applies non-perturbatively.)  Our first job is to clarify the role of the gravity partition function \eq{bulk}.

\paragraph{4.2.2.1~~The gravity partition function.}  To clarify the correspondence between the bulk path integrals and the formalism in Section \ref{duality}, let us recall the usual correspondence between the path integral quantization of quantum mechanics and the canonical formalism.

In quantum mechanics, the path integral with prescribed boundary conditions is the transition amplitude between the two endpoints, satisfying the Schr\"odinger equation:
\bea\label{QM}
\bra q_2,t_2|q_1,t_1\ket=\int_{q_1=q(t_1)}^{q_2=q(t_2)}{\cal D}q(t)~\exp\left(i\int_{t_1}^{t_2}\dd t\,L(q,\dot q)\right)~,
\eea
with the measure defined as the limit of a summation over discrete path segments, properly (re-)normalised to 1. For Lorentzian quantum field theories, and under standard assumptions about the vacuum state, one deforms the integration contour in the complex plane in such a way that highly excited contributions are suppressed in the large-time limit.
For instance, it is a standard result that in the Heisenberg picture the two-point function of a scalar field theory is given by:
\bea\label{corrf}
\bra0|\,T\left(\hat\phi(x_1)\hat\phi(x_2)\right)|0\ket=\lim_{T\rightarrow\infty(1-i\epsilon)}{\int{\cal D}\phi~\phi(x_1)\, \phi(x_2)~\exp\left[i\int_{-T}^T\dd^4x~{\cal L}\right] \over\int{\cal D}\phi~\exp\left[i\int_{-T}^T\dd^4x~{\cal L}\right]}~.
\eea
Functional methods can be adapted to work for both abelian and non-abelian quantum field theories. (In fact, there are several important respects in which the funcitonal formalism is more useful than the canonical formalism: cf.~e.g. Weinberg (1995), p. 377). In particular, correlation functions such as \eq{corrf} can be calculated, once the generating functional is known, by functional differentiation with respect to the appropriate source. 

Our aim here is not to identify the correlator \eq{corrf} with the gravity partition function \eq{bulk} but to compare \eq{corrf} with the {\it boundary}  two-point functions \eq{2pt}, which are the quantities of interest. The latter are {\em not} calculating correlation functions of {\it bulk} operators; instead, \eq{bulk} defines {\it boundary} correlation functions, such as \eq{2pt}, as functional derivatives with respect to the asymptotic {\it boundary values} of (boundary conditions on) the bulk fields. The functional differentiation with respect to the boundary value of the field gives the one-point function of the canonical momentum, $\bra\Pi_\phi(x)\ket$, associated with that field (at the boundary!) (see pp. 3-4 and 10 of Witten (2001)). Thus, the correlation functions \eq{2pt}, calculated on either side of the duality, are the analogs of the expressions \eq{corrf} in ordinary functional methods in quantum field theory. 

Of course, this interpretation is somewhat formal, because the bulk theory is Euclidean; but it makes sense in a Euclidean theory to consider conjugate pairs of canonical variables $(\phi_{(0)},\Pi_{\phi})$. The state $|\phi_{(0)}\ket$ is then, in a way analogous to $|q_1,t_1\ket$ in \eq{QM}, the state of the bulk field with asymptotic boundary condition $\phi_{(0)}(x)$. This procedure thus defines a {\it boundary} Hilbert space, as is appropriate in a diffeomorphism invariant theory.

Under this interpretation of \eq{bulk} as generating the correlation functions \eq{2pt} in the boundary Hilbert space of the bulk theory, the path integral itself \eq{bulk}, with $\phi_{(0)}=0$, is the vacuum wave functional: $Z[0]=\bra0|0\ket$. Here, $|0\ket$ is the state with standard, conformally invariant, boundary conditions $\phi_{(0)}=0$. The functional derivatives of $Z[\phi_{(0)}]$, setting the source to zero, are the boundary correlation functions of the canonical momenta, $\bra\Pi_{\phi}(x_1)\ldots\Pi_{\phi}(x_n)\ket$, evaluated in the vacuum state.

\paragraph{4.2.2.2~~Calculating the quantities.}  We now show how to calculate the quantities. 
For a fixed AdS background, the Einstein equations for a scalar field with fixed boundary condition $\phi_{(0)}$ at infinity have a unique regular solution:
\bea\label{bulkf}
\phi(r,x)=\int\dd^dy~{r^{\Delta}\over(r^2+|x-y|^2)^{\Delta}}~\phi_{(0)}(y)~,
\eea
where $\Delta$ is defined as in \eq{dim}. The factor raised to the power $\Delta$ is called the `bulk-to-boundary' propagator. This result can be used to calculate the expectation value of the boundary operator (the canonical momentum associated to the boundary value of the field) in the semi-classical limit.\footnote{We call this limit `semi-classical' because we have in mind the effective field theory approach discussed in Section \ref{dicty}, where it is possible to calculate quantum corrections: see \eq{QEA}.} This is done by plugging the solution into the leading classical approximation to \eq{bulk}, which is just the action evaluated on solutions \eq{bulkf} that satisfy the prescribed boundary conditions. By functional differentiation of \eq{adscft} one finds:
\bea\label{1pt}
\bra\Pi_\phi(x)\ket_{\phi_{(0)}}=\int\dd^dy~{\phi_{(0)}(y)\over|x-y|^{2\Delta}}~,
\eea
which is evaluated on the state $|\phi_{(0)}\ket$. Of course, setting $\phi_{(0)}=0$ here gives a zero result---from the bulk theory, a consequence of regularity; also a consequence of translation invariance, from the boundary point of view. The two-point function naturally gives a non-zero result; and it reproduces \eq{2pt}, given earlier as  the CFT result.\\

We have so far discussed scalar fields. What about the Hilbert space of the gravitational field? In a similar way, one can fix the boundary value of the metric at infinity and solve the bulk equations similarly to \eq{bulkf}. Again, there is a canonical momentum operator $\Pi_{ij}(x)$ associated with the boundary value of the metric, whose expectation value agrees with the boundary stress-energy tensor leading to the anomaly \eq{anomaly}; and the latter was, as we mentioned, a non-perturbative result. In fact, the correspondence between the canonical momentum and the boundary stress-energy tensor has been shown to hold to arbitrarily high order in perturbation theory. Thus Faulkner et al. (2014) contains a linearised result, but with arbitrarily high order curvature corrections; and  Sen and Sinha (2014)  contains a non-linear result. The significance of these higher-curvature corrections is that they are sub-leading in the $1/N$ expansion, i.e.~they are quantum (stringy) corrections to the leading supergravity result, as we also explained after \eq{QEA}. 

We have argued that the bulk path integral \eq{bulk} is equivalent to the canonical quantization of the bulk field $\phi(r,x)$. However, we also stressed that it would be a mistake to think of \eq{bulk} as calculating correlation functions of the type \eq{corrf}; instead, the field has to be pushed to the boundary, as in \eq{bulkf}; and both the Hilbert space and the operators are defined on the boundary. This is what one expects in a diffeomorphism invariant theory with no local degrees of freedom and no local quantities in the bulk. The quantities are either non-local, or localized on boundaries.

\subsubsection{An objection and a reply}\label{obj}

The discussion in Section \ref{matches}---especially the clarification that \eq{bulk} does not give correlation functions for bulk operators in direct analogy with functional methods in a quantum field theory---prompts an objection to what we have done. Thus one might say that what we have done is trivial: we {\it assumed} the correspondence \eq{adscft} and repackaged the bulk quantities in terms of the boundary quantities; so that no independent Hilbert space for the bulk theory was provided. What we have done is trivial because---the objection continues---we know, from diffeomorphism invariance, that the bulk Hilbert space and all local operators must be associated with the boundary. Further, the asymptotic symmetries (the $d$-dimensional conformal group; or, in fact, the superconformal group) also require that the Hilbert space and its operators must be representations of the (super-) conformal group in one fewer dimension.\footnote{Actually, up to anomalies!} So something like \eq{adscft} was to be trivially expected.

There are two things to say in reply to this objection.

(i) The first is that: although the Hilbert space and the operators obtained above are the same between the bulk and the boundary (they indeed fall in representations of the superconformal group), the procedures leading to their definitions are completely independent from each other. For in one case, the Hilbert space is constructed from CFT quantities; in the other, one looks at the asymptotics of bulk fields \eq{bulkf} and their canonically conjugate variables. In particular, the bulk side of this story finds its own justification in the diffeomorphism invariance of the theory, which forces us to define a {\it boundary} Hilbert space with a representation of the conformal group.

(ii) The second reply is that: although the matching of the Hilbert spaces, for weakly coupled gravity, might be more or less trivial (with the hindsight of AdS/CFT!) and dictated by conformal invariance---recall that the CFT is strongly coupled. And this means that the non-triviality (physicists would say: the dynamical information!) lies in the quantum corrections: which also match. For instance, the one non-trivial number that needs to be compared between the bulk two-point function obtained from \eq{1pt} and the boundary two-point function \eq{2pt} is the anomalous dimension $\Delta$; and, though its leading behaviour again follows from conformal invariance alone, its quantum corrections are highly non-trivial and have been found to match on the two sides, in those special cases where the two sides can be compared. Something similar holds for the other examples we gave earlier: the anomaly \eq{anomaly} and the match between the full stress-energy tensors; as well as the non-perturbative matching of SL$(2,\mathbb{Z})$ invariant couplings, mentioned earlier (Section \ref{dicty}, (b)).\\
  
To sum up this Subsection: the formal structure of AdS/CFT duality exemplifies the formulation of duality in Section \ref{duality}. That is:\\
\indent  (1): The bulk side  gives rise to a theory $T_{\tn{bulk}} = \langle \mathcal{H}_1, \mathcal{Q}_1 \rangle$, while the boundary dual gives rise to $T_{\tn{bdy}} = \langle \mathcal{H}_2, \mathcal{Q}_2 \rangle$; where $\mathcal{H}_i$ is a Hilbert space, $\mathcal{Q}_i$ is some algebra of operators, and unitary dynamics is implicit.\\
\indent (2): Explicit results in perturbation theory, as well as some extant non-perturbative results, confirm that the definition of duality is instantiated; and is in fact much more insensitive to quantum corrections than one might have hoped for. In other words: various results strongly suggest that \eq{adscft} yields  duality maps $d_s$ and $d_q$, which are isomorphisms of Hilbert spaces and operators respectively, and which satisfy \eq{obv1} and \eq{obv2}. 

\section{Some complications for gauge invariance}\label{invarces}

\subsection{Gauge invariance and duality in AdS/CFT}\label{GIduality}

Let us now turn to the topic of gauge invariance in AdS/CFT. One typically identifies gauge symmetry, in the sense (Local) of Section 2, at the level of the classical Lagrangian: which enters into the path integral formulae of \eq{adscft}. In the bulk theory, the main (Local) gauge symmetry is diffeomorphism symmetry. Furthermore, one can add a Yang-Mills type (Local) gauge symmetry to the bulk by coupling a gauge field $A \in \Omega^1 (M) \otimes \mathfrak{u}(1)$ to the bulk metric, where $\Omega^1 (M)$ denotes the space of $1$-forms on the bulk spacetime $M$. In the boundary CFT, on the other hand, the (Local) gauge symmetry is specified in terms of $\mbox{SU}(N)$ gauge transformations acting on the boundary gauge fields $A \in \Omega^1 ( \partial M ) \otimes \mathfrak{su}(N)$, where $\partial M$ denotes the boundary spacetime.

What about gauge invariance at the quantum level? Were we dealing (albeit perturbatively) with a simple case of gauge theory (e.g.~QED), we would obtain the \textit{quantum} state space by constructing a nilpotent Hermitian (BRST) operator $Q$ and defining the physical (gauge-invariant) states as those annihilated by $Q$: cohomology then ensures that each state represents only a gauge orbit.\footnote{Failing to perform this move, i.e.~naively quantizing the gauge fields of the theory, would lead to an indefinite Fock space (i.e.~a space containing states with negative norm).} Thus, for two such quantum theories $T_1$ and $T_2$, there can be no question about whether a duality map between them relates their classical gauge symmetries: these symmetries are simply not represented to begin with. 
 
In contrast, one cannot proceed in this way for the gauge theories involved in AdS/CFT (and many other dualities) because each side of the duality typically contains a non-perturbative sector, for which we cannot directly construct the quantum state space. It is precisely here that the duality relation \eq{adscft} is strikingly useful: e.g., by exchanging the non-perturbative sector of $T_{\tn{bdy}}$ with the perturbative sector of $T_{\tn{bulk}}$ (i.e.~semi-classical supergravity), it allows us to indirectly construct the quantum state space of $T_{\tn{bdy}}$ by means of $T_{\tn{bulk}}$. 

But this also opens up a Pandora's box for gauge invariance! For: prima facie, it allows that the (Local) gauge symmetries of $T_{\tn{bulk}}$ might be related to the symmetries of $T_{\tn{bdy}}$. As we will now explain, Horowitz and Polchinski have argued against this possibility, i.e.~for what Section \ref{samedifferent}  called `invisibility'. 
 
 Recall that (naively) a duality is a bijection between the gauge-invariant contents (states and quantities) of two theories $T_1$ and $T_2$.  Now suppose one is sceptical that a duality would  match elements of a gauge orbit $G_1$ ($G_1 \subset {\cal S}_1$) in theory  $T_1$ with elements of the dual gauge orbit $G_2 := d_s(G_1)$ in $T_2$.  That is, in terms of Section \ref{samedifferent}'s everyday analogy with translations between languages : one doubts that a translation will match each synonym in a set of synonyms in $T_1$  with a synonymous member of an equinumerous set of synonyms in $T_2$.  Then one would naturally expect that the  the duality mapping (e.g.~as given for AdS by \eq{adscft}) secures the following: 
  \begin{quote}
  (Invisibility) Each gauge symmetry of $T_1$, i.e.~permutation of ${\cal S}_1$ that leaves each gauge-orbit invariant (in the analogy: permutation of $T_1$'s words that leaves each synonymy equivalence class invariant)  carries over to  only the identity permutation on ${\cal S}_2$; and vice versa. In this sense, the duality \textit{only} relates the gauge orbits of $T_1$ and $T_2$. 
    \end{quote}

This is indeed the definition of `invisibility' that Horowitz and Polchinski use in (2006, Section 1.3.2). In (2015, Section 2.3 and 2.4), Polchinski demonstrates this property for the duality between $p$-form gauge theories; and in (2006, Section 1.3.2), it is claimed that in AdS/CFT, even the (Local) gauge symmetry of bulk, i.e.~diffeomorphism symmetry, displays such `invisibility', thus showing that diffeomorphism symmetry (and relatedly, spacetime) is an `emergent' property: ``the gauge variables of AdS/CFT are trivially invariant under the bulk diffeomorphisms, which are entirely invisible in the gauge theory'' (Horowitz and Polchinski (2006:~p.~12)). 

Is this latter claim correct? To be sure, there exist simple examples of `invisibility' in AdS/CFT, from both the bulk and the boundary perspective. From the bulk: since all correlation functions defined by \eq{adscft} are invariant under boundary (Local) gauge symmetry, this symmetry is not seen in the bulk. We saw this explicitly in the construction of the operator of dimension $\D=4$ in SYM, corresponding to a massless scalar field in the bulk, in Section \ref{dicty}. From the boundary: The AdS$_4$ example at the end of Section \ref{dicty} gives an example of how a U(1) invariant combination of the bulk field (the electric field) is related to a U(1) invariant boundary quantity (a conserved current). Here the U(1) symmetries in the bulk and in the boundary  are completely unrelated.

Another example of invisibility concerns the reparametrisation symmetry of a string in AdS$_5\times S^5$. Consider a string whose two-dimensional world-sheet $\Sigma$ intersects the AdS$_5$ boundary along a circle ${\cal C}$. The path integral over the configuration of this string with given boundary condition ${\cal C}$ contains in its integrand the Polyakov action for the string. Among the path integration variables are, in particular, all the string world-sheet geometries with the given boundary condition. This theory has a reparametrisation symmetry with respect to the coordinates parametrising the world-sheet. Now consider the dual of this in ${\cal N}=4$ SYM: it is the expectation value of a Wilson loop around the contour ${\cal C}$, which describes a quark-antiquark pair. The result in SYM theory is that the expectation value of this Wilson loop decays exponentially with the quark-antiquark potential (the Coulomb potential). In this example, the duality is such that the semi-classical approximation to the Polyakov path integral in the bulk gives the quark-antiquark potential of the boundary Wilson loop, evaluated at strong coupling. The AdS$_5$ calculation amounts to minimising the surface area ending on the contour ${\cal C}$ and evaluating the Nambu-Goto action for that classical classical string configuration. On the SYM side, the Wilson loop is constructed purely out of the gauge theory variables---gauge fields and scalars---and there is clearly no fundamental string present, hence no world-sheet. In the bulk theory, there is the reparametrisation symmetry of the Polyakov action; and this is invisible to the SYM theory---hence we have here another example of an invisible symmetry.

Based on these simple examples, one might expect that the main (Local) gauge symmetry of the bulk theory, viz.~diffeomorphism symmetry, is also entirely invisible in the boundary theory. In the next Section, we shall argue that this claim is false. Roughly speaking:\\
\indent (i) not all diffeomorphisms are invisible in AdS/CFT; and furthermore:\\
\indent (ii) there \textit{is} a sense in which the duality map \eq{adscft} maps (Local) gauge symmetries in the bulk to position-dependent (Redundant) symmetries in the boundary.

\subsection{AdS/CFT's visible and invisible diffeomorphisms}\label{noinvis}
 
As we announced in Section \ref{GIduality}, we will now distinguish the diffeomorphisms of the bulk theory that are {\it invisible}  in the boundary theory from those that are {\it visible}. The analysis will also give us a precise characterisation of these two classes. 

We construe the notion of `visible' diffeomorphism, along the lines of Horowitz and Polchinski's own examples, as one that does not restrict to the identity map on the boundary: because, when it does not tend to the identity map, the diffeomorphism can be `seen' by the boundary theory: the fields in the boundary theory are modified by it.  `Invisible' means that the diffeomorphism does restrict to the identity. Thus, in specifying the classes of diffeomorphisms that we call `visible' and `invisible', we will make use of the following three conditions: \\
\\
(i) ~~(Fixed): the diffeomorphisms leave the form of the bulk metric (\eq{normalp} below) fixed;\footnote{Notice that \eq{normalp} is a form, i.e. a class, of metrics so that fixing \eq{normalp}  does not require being an isometry. See De Haro (2016) for a full account.} \\
\\
(ii) ~(Invisibility): this condition has two parts, roughly: invariance of the fields and triviality of the diffeomorphisms. Spelling this out, the diffeomorphisms: \\
\indent(iia)~(Invariance): leave all boundary quantities invariant;\\
\indent(iib)~are equal to the identity, at the boundary;\\
\\
(iii) (Existence): the diffeomorphisms are non-trivial in the bulk. \\
\\
The large-distance behaviour of AdS/CFT is qualitatively different in even and odd boundary dimensions. The reason for this is the conformal anomaly, which only appears for even $d$. Therefore, at this point we must treat these two cases separately.

\subsubsection{Odd $d$}\label{oddd}

It can be shown that, if the boundary dimension $d$ is odd, then there are {\it no} diffeomorphisms that satisfy the conditions (i), (ii), (iii) in conjunction. More precisely: under the assumption that the form of the bulk metric is fixed, there are no diffeomorphisms that are equal to the identity at infinity and extend non-trivially to the bulk. 

We shall sketch the argument for this in a moment. But to show that there {\it are} nontrivial visible diffeomorphisms, we need to consider relaxing the above conditions: this relaxation will also lead in to our final point, about (Local) diffeomorphisms that are (Redundant). 

We get {\it invisible diffeomorphisms} if we drop condition (i). Any diffeomorphism satisfying (ii) and (iii) will be invisible, and will be a candidate for a discussion of emergence along the lines of Horowitz and Polchinski (2006). But beware that not {\it all} bulk diffeomorphisms are invisible. We can find an important class that is visible, precisly by replaceing (Invisibility) by the weaker:
\\
\\
(ii') (Invariance): that is, (iib) is dropped. The bulk diffeomorphisms can be non-vanishing  on the boundary, but must  leave all the CFT quantities invariant.\\
We will see that this condition gives us, in a precise manner, the connection between the asymptotic diffeomorphisms of the bulk and the boundary conformal group. \\

A sketch of the argument that shows the incompatibility of (i)-(iii) is as follows. Let the bulk metric be that of an Einstein space (i.e.~a solution of general relativity's field equations with a negative cosmological constant), so that an {\it arbitrary} metric is induced at the boundary. Let us call this metric $g_{(0)ij}(x)$. A remarkable result by Fefferman et al.~(1985: cf. also their 2012) shows that, for a space that satisfies Einstein's equations with a negative cosmological constant, and given a conformal metric at infinity, the line element can be written in the following form:
\bea\label{normalp}
\dd s^2={\ell^2\over r^2}\left(\dd r^2+g_{ij}(r,x)\,\dd x^i\dd x^j\right).
\eea
Here, as before, $r$ parametrises the distance to the boundary, and the boundary is parametrised by the coordinates $x^i (i=1,\ldots,d)$. At $r=0$, $g_{ij}(r,x)$ is a real analytic function of $x$, and induces a conformal metric on the boundary: $g_{(0)}(x):=g_{ij}(0,x)$. Clearly, the above is a generalisation of \eq{AdSmetric}, which assumed a flat 4D boundary metric. For now the boundary metric (i.e. the representative of the conformal class) is only assumed to be any real analytic function of $x^i$. (De Haro et al (2016: Section 6.1.1.) gives more details.)

The condition (Fixed) gives the following transformation property for $g_{ij}(r,x)$:
\bea\label{dgij}
\d g_{ij}(r,x)=\nabla_i\xi_j(r,x)+\na_j\xi_i(r,x)+2\xi(x)\left(1-2r\,\pa_r\right)g_{ij}(r,x)~.
\eea
Here, $(\xi,\xi^i)$ is an infinitesimal vector field\footnote{We present the argument in its infinitesimal form because it is simpler. However, the argument readily generalises to the finite form of the diffeomorphisms.} that parametrises the infinitesimal diffeomorphisms in $(r,x^i)$. Indeed, one easily recognises, in the first two terms, the usual transformation property of a 2-tensor; the last term comes from the radial diffeomorphisms, and can be thought of as a conformal transformation.

Next, (Invisibility). It consists of two requirements: (iia) (Invariance), in particular of the boundary metric; and also: (iib) the diffeomorphisms vanish at the boundary. Requiring (iia) means $\d g_{(0)ij}=0$, which is achieved by setting $r=0$ in \eq{dgij}. That condition leads to the following equation for the (now visible) boundary diffeomorphisms:
\bea\label{conf}
\nabla_i\xi_j(x)+\nabla_j\xi_i(x)-{2\over d}\,g_{ij}(x)\,\nabla^k\xi_k(x)=0~.
\eea
There are two important points about this equation, which is the mathematical representation of (Invariance):\\
\indent (a) This equation is the condition for an infinitesimal {\it boundary} coordinate transformation $\xi_i$ to give a local scale transformation: thus the $\xi_i$ generate the {\it boundary conformal group}.\footnote{To the best of our knowledge, the present derivation, and the characterisation of the conformal group as arising from the asymptotic form of the diffeomorphisms satisfying (iia), for {\it arbitrary} boundary metrics, has not appeared in the literature. In particular, the diffeomorphisms we consider here are more general than those considered in Imbimbo et al.~(2000: Eqs.~(2.6)-(2.9)), Skenderis~(2001: Eq.~(8)). These authors do not obtain the condition for the boundary diffeomorphisms to be conformal, Eq.~\eq{conf}. See De Haro~(2016). } Diffeomorphisms satisfying (iia) thus must solve \eq{conf}, and they form the class of {\it visible} diffeomorphisms.\\
\indent (b) As a consequence of (a), when one requires in addition to (Invariance), also  (Invisibility), i.e.~that the boundary coordinate transformations vanish, i.e.~$\xi_i=0$: then one can subsequently show that the diffeomorphism in fact is identically zero throughout the bulk. Requiring (Invisibility) of the diffeomorphism thus deprives it of  (Existence).

\subsubsection{Even $d$}


There are two reasons why we restricted our comments to the case of odd $d$.\footnote{Notice that the case of odd $d$ includes the physically interesting case of an 11-dimensional bulk that, when seven dimensions are compactified, gives rise to a theory of gravity in four dimensions, with a 3-dimensional CFT dual. This case is physically interesting because the four-dimensional gravity descends from an 11-dimensional theory, so-called M-theory, which is conjectured to have well-defined quantum behaviour. See e.g.~Green (1999: Section 1), Maldacena (1998: Section 3).}

 (i): In the case that $d$ is even, the diffeomorphisms \eq{conf} are not invariances of the theory. Due to the conformal anomaly \eq{anomaly}, such putative diffeomorphisms in fact violate (Invariance). It can be shown that the breaking of conformal invariance is tied up with a breaking of diffeomorphism invariance: the holographic stress-energy tensor $\bra T_{ij}\ket$ no longer transforms as a rank-two tensor. 
 
 (ii): The derivation sketched in Section \ref{oddd} assumed a regular expansion of the boundary metric $g_{ij}(r,x)$ near $r=0$. When $d$ is even, such expansion would not give a solution of Einstein's equations, except in very special cases. To solve Einstein's equations for even $d$, one needs to include  in the expansion terms that are logarithmic in $r$. These logarithmic terms are responsible for the conformal anomaly. 


Recall our discussion, in Section \ref{motivating} (1)-(3), of active versus passive symmetries. The diffeomorphisms considered in this section are construed as active, which is also the most interesting interpretation when discussing the hole argument (see Section \ref{gauge}). However, similar considerations apply to passive diffeomorphisms (De Haro~(2016)). 

\section{Galileo's ship, (Local) vs.~(Redundant)}\label{gauge}

We conclude with two points about the asymptotic bulk symmetries that we defined as having the property (Invariance) but not (Invisibility): the first is technical/historical, and the second is interpretive.

First, the identification between asymptotically AdS bulk symmetries and boundary conformal transformations pre-dates the AdS/CFT correspondence and can be seen as one of its early intimations: working only within the context of classical 3D gravity with negative cosmological constant, Brown and Henneaux (1986) showed that diffeomorphisms preserving the AdS$_3$ boundary conditions $g_{(0)ij}=\eta_{ij}$ ($i=1,2,3$) (the three-dimensional Lorentz metric) give rise to (two copies of) the Virasoro algebra in two dimensions.
Furthermore, they showed that when such asymptotic symmetries are not isometries of the spacetime (as they would be if the spacetime were globally AdS$_3$) then the symmetry algebra acquires a central extension (i.e.~the brackets of the generators no longer close, but acquire an additional term referred to as a `central charge'). The AdS/CFT correspondence extends this result in higher dimensions; the central charge is then generalised as the conformal anomaly (see \eq{anomaly}), and it also appears as the normalization of the two-point function of the stress-energy tensor of the boundary theory (see e.g.~Sections 2.1 and 2.2 of Gubser et al.~(1998)). 

Second, how should we interpret these asymptotic bulk diffeomorphisms? One might be inclined to diagnose an interpretive problem here if one naively took these bulk diffeomorphisms to \textit{necessarily} be (Redundant) transformations of the bulk theory, thus yielding an interpretive conflict with their boundary `translation', viz.~the boundary (global) conformal transformations, which are potentially non-(Redundant) when restricted to subsystems.\footnote{When restricted to a proper subsystem, these global transformations---like any other global symmetries---can be used to create a relational difference in states between the subsystem and its environment.}

But this would be a mistake. As we mentioned at the start of Section \ref{samedifferent}, and as discussed in both the philosophical and physical literature, not all (Local) gauge transformations are (Redundant). In the philosophical literature, this topic has been discussed under the head of whether gauge symmetries possess the sort of `direct empirical significance' that is characteristic of global symmetries. The relevant notion of `direct empirical significance' comes from Galileo's famous `ship thought experiment', wherein a global (spacetime, in this case) symmetry is used to create a relational physical difference between a proper subsystem and an environment \textit{when} the (action of the) symmetry is restricted to the subsystem. The mistaken diagnosis arises only if one denies that gauge symmetry can be used to construct an analogous Galileo's Ship scenario.

 However, as Teh (2015) and Greaves and Wallace (2014) argue, by requiring that (Local) gauge transformations preserve appropriate asymptotic boundary conditions, one can construct an asymptotic symmetry group that is perfectly capable of displaying Galileo's Ship type phenomena (roughly: one quotients the boundary-condition-preserving group of symmetries by its subgroup of asymptotically trivial symmetries). The observation that asymptotically AdS bulk diffeomorphisms discussed above are a special case of this phenomenon thus removes the putative obstacle to matching the interpretation of the asymptotic bulk diffeomorphisms and the boundary global conformation transformations (it is for this reason that Brown and Henneaux (1986) also refer to such asymptotic symmetries as `global' symmetries). 

We can now establish the connection with Einstein's hole argument, promised at the end of Section 2. For our analysis reveals a relevant class of diffeomorphisms for which an analogue of the hole argument can be formulated in the context of AdS/CFT, namely the invisible diffeomorphisms: they tend to the unit map asymptotically, but are non-trivial in the bulk. So here, the bulk is the analogue of Einstein's hole. Such a diffeomorphism, being a symmetry of the theory, identifies points $p,q$ in spacetime that one naively considers different (because labelled by different coordinates $\psi_1(p),\psi_2(q)\in\mathbb{R}^{d+1}$): for any two points $p\in U_1,q\in U_2$ in neighbourhoods $U_1,U_2$ contained in the hole, $p\sim q$ iff there is an invisible diffeomorphism $\f:U_1\rightarrow U_2$ such that: $q=\f(p)$ (and the metric, and other physical quantities, are simultaneously pulled back). As we have argued, such diffeomorphisms are restricted, but exist: the relevant class is that of diffeomorphisms that are (Invisible) but not (Fixed).  

Finally, notice that the above analysis still leaves open the question of how the asymptotic bulk diffeomorphisms (and their corresponding boundary conformal transformations) should be interpreted. If we are working in the context of theories of the universe and we adopt the `internal point of view' (see Section \ref{defth} for more on both notions),\footnote{The question on the necessity of a theory of the universe to support the `internal point of view' is discussed in section 2.4.1 of De Haro (2015).} then two states related by such transformations are physically equivalent: the transformations count as (Redundant) gauge symmetries. On the other hand, the contrary `external point of view' interprets \eq{adscft} as coupled to an already interpreted physical background, represented by $\phi_{(0)}(x)$ (and the fixed metric $g_{(0)ij}$); and in that case, the diffeomorphisms would not be (Redundant) but would become physical. When we describe Galileo's ship relative to the quay, its state of motion is indeed physically meaningful!

To conclude, we have investigated various aspects of the relationship between gauge symmetries and dualities, in particular the way in which gauge symmetries behave under the holographic duality map. Our results show that various familiar concepts---``gauge'', ``symmetry'', and ``diffeomorphism invariance''---are both complicated and enriched when analyzed within the setting of dualities.

\section*{Acknowledgements}
\addcontentsline{toc}{section}{Acknowledgements}
We thank Kostas Skenderis and two anonymous referees for valuable comments. We also thank the John Templeton Foundation for supporting this research, through grant 28929.

\section*{References}
\addcontentsline{toc}{section}{References}

\hspace{3ex}  Ammon, M., Erdmenger, J.~(2015). ``Gauge/Gravity Duality. Foundations and Applications''. Cambridge University Press, Cambridge.\\

Barrett, T. and Halvorson, H. (2015), `Glymour and Quine on theoretical equivalence', Pittsburgh archive: http://philsci-archive.pitt.edu/11341.\\

Brown, J.~D.~and Henneaux, M~(1986). ``Central Charges in the Canonical Realization of Asymptotic Symmetries: An Example from Three-Dimensional Gravity'',{\it Communications in Mathematical Physics} {\bf 104} (1986) 207.\\

Butterfield, J. and Bouatta, N. (2015). `On emergence in gauge theories at the {'t} Hooft limit', European Journal for Philosophy of Science 5, 2015, 55-87.\\

De Haro, S., Skenderis, K., and Solodukhin, S. (2001). ``Holographic reconstruction of spacetime and renormalization in the AdS/CFT correspondence", \emph{Communications in Mathematical Physics}, 217(3), 595-622. [hep-th/0002230].\\

De Haro, S. (2015), `Dualities and Emergent Gravity: Gauge/Gravity Duality', {\em Studies in History and Philosophy of Modern Physics}: this volume.\\

De Haro, S., Mayerson D. and Butterfield J.~(2016). `Conceptual aspects of Gauge/gravity duality', forthcoming in {\em Foundations of Physics}.\\

De Haro, S. (2016). ``Is Diffeomorphism Invariance Emergent?'', in preparation.\\

Deser, S.~, Schwimmer, A.~(1993). ``Geometric classification of conformal anomalies in arbitrary dimensions,''
  {\it Physics Letters B}, {\bf 309} 279
  [hep-th/9302047].\\

Dieks, D., Dongen, J. van, Haro, S. de, (2015). ``Emergence in Holographic Scenarios for Gravity'', PhilSci 10606, arXiv:1501.04278 [hep-th]. Forthcoming in {\it Studies in History and Philosophy of Modern Physics}.\\

Faulkner, T., Guica, M., Hartman, T., Myers, R.C., Van Raamsdonk, M. (2014), `Gravitation from Entanglement in Holographic CFTs',  {\em Journal of High Energy Physics} {\bf 1403} 051  [arXiv:1312.7856 [hep-th]].\\

Fefferman, C.~and Graham, C.R.~(1985). ``Conformal Invariants'', in {\it Elie Cartan et les Math\'ematiques d'aujourd'hui}, Ast\'erisque, 95.\\

Fefferman, C.~and Graham, C.R.~(2012). ``The Ambient Metric'', Annals of Mathematics Studies,
number 178. Princeton University Press: Princeton and Oxford; (and at: http://arxiv.org/abs/0710.0919). \\

Greaves, H. and Wallace, D. (2014), `Empirical Consequences of Symmetries', {\em British Journal for the Philosophy of Science} {\bf 65} pp. 59-89.\\

Green, M.~B.~(1999). ``Interconnections between type II superstrings, M theory and N=4 supersymmetric Yang-Mills,''
  {\it Lecture Notes in Physics}, {\bf 525} 22
  [hep-th/9903124].\\

Halvorson, H. (2015), `Scientific theories', forthcoming in {\em The Oxford Handbook of Philosophy of Science}, Pittsburgh archive: http://philsci-archive.pitt.edu/11347/.\\

Horowitz, G and Polchinski, J. (2006), `Gauge/gravity duality', in {\em Towards quantum
gravity?}, ed. Daniele Oriti, Cambridge University Press; arXiv:gr-qc/0602037. \\

Huggett, N (2015). ``Target Space $\neq$ Space'', {\em Studies in History and Philosophy of Modern Physics}: this volume.\\

Imbimbo, C., A.~Schwimmer, S.~Theisen and S.~Yankielowicz~(2000). ``Diffeomorphisms and holographic anomalies,''   {\it Classical and Quantum Gravity} {\bf 17} 1129
  doi:10.1088/0264-9381/17/5/322
  [hep-th/9910267].\\

McGreevy, J. (2010). ``Holographic duality with a view to many-body physics''. {\em Advances in High Energy Physics}; arXiv:hep-th/0909.0518.\\

Maldacena, J. (1998). ``The large $N$ limit of superconformal field theories and supergravity'', {\em Advances in Theoretical and Mathematical Physics} {\bf 2}, p. 231: 
arxiv: hep-th/9711200.\\

Matsubara, K.~(2013). ``Realism, underdetermination and string theory dualities." {\it Synthese}, 190.3: 471-489.\\

McKenzie, K. (2015), `Relativities of Fundamentality', submitted to {\em Studies in History and Philosophy of Modern Physics}\\

Polchinski, J.~(2015), ``Dualities of Fields and Strings,'' forthcoming in {\em Studies in History and Philosophy of Modern Physics};    
arXiv:hep-th/1412.5704. Contribution to this volume.\\

Rickles, D. (2011). `A philosopher looks at string dualities', {\em Studies in History and Philosophy of Modern Physics}, {\bf 42} (1), 54-67.\\

Rickles, D. (2012). `AdS/CFT duality and the emergence of spacetime', {\em Studies in History and Philosophy of Modern Physics}, {\bf 44} (3), 312-320.\\

Rickles, D. (2015). ``Dual theories: `same but different' or `different but same'?'', {\em Studies in History and Philosophy of Modern Physics}: this volume.\\

Sen, K., Sinha, A. (2014), `Holographic stress tensor at finite coupling', {\em Journal of High Energy Physics} {\bf 1407} 098  [arXiv:1405.7862 [hep-th]].\\

Skenderis, K.~(2001). ``Asymptotically Anti-de Sitter space-times and their stress energy tensor,'' {\it International Journal of Modern Physics} A {\bf 16} 740
  doi:10.1142/ S0217751X0100386X
  [hep-th/0010138].\\

Teh, N.J. (2013). `Holography and emergence', {\em Studies in History and Philosophy of Modern Physics}, {\bf 44} (3), 300-311.\\

Teh, N. (2015), `Galileo's Gauge: Understanding the Empirical Significance of Gauge Symmetry', forthcoming in {\em Philosophy of Science}. \\

Teh, N. and Tsementzis, D. (2015), `Theoretical Equivalence in Classical Mechanics and its
relationship to Duality', {\em Studies in History and Philosophy of Modern Physics}: this volume.\\

Weinberg, S.~(1995). ``The Quantum Theory of Fields'', volume I. Cambridge University Press (Cambridge and NY). \\

Witten, E.~(1995), `String theory dynamics in various dimensions',  {\em Nuclear Physics B} {\bf 443}, 85   [hep-th/9503124].\\

Witten, E.~(2001). ``Multitrace operators, boundary conditions, and AdS / CFT correspondence,''
  hep-th/0112258.

\end{document}